\documentclass[fleqn,usenatbib]{mnras}
\usepackage[T1]{fontenc}

\DeclareRobustCommand{\VAN}[3]{#2}
\let\VANthebibliography\thebibliography
\def\thebibliography{\DeclareRobustCommand{\VAN}[3]{##3}\VANthebibliography}

\usepackage{graphicx}
\usepackage{amsmath}	% Advanced maths commands
\usepackage{amssymb}	% Extra maths symbols
\usepackage{bm}
\usepackage{hyperref}
\usepackage[normalem]{ulem}
\usepackage{multirow}

\usepackage{newtxtext}
\usepackage[varvw]{newtxmath}
\let\vec\bm% or \mathbf

\newcommand{\diff}{\mathrm{d}}
\newcommand{\e}{\mathrm{e}}
\newcommand{\I}{\mathrm{i}}

\def\vx{\vec{x}}

\def\vk{\vec{k}}

\newcommand{\dc}{\delta_\mathrm{c}}
\newcommand{\fec}{f_\mathrm{ec}}
\newcommand{\rhoM}{\rho_\mathrm{m}}
\newcommand{\rhoL}{\rho_\Lambda}

\title[The simplest excursion set theory]{Accurate halo mass functions from the simplest excursion set theory}

\author[M. S. Delos]{
M. Sten Delos\thanks{E-mail: mdelos@carnegiescience.edu}
\\
The Observatories of the Carnegie Institution for Science, 813 Santa Barbara Street, Pasadena, CA 91101, USA
\\
Max Planck Institute for Astrophysics, Karl-Schwarzschild-Str. 1, 85748 Garching, Germany
}

\date{Accepted XXX. Received YYY; in original form ZZZ}

\pubyear{2023}

\begin{document}

\label{firstpage}
\pagerange{\pageref{firstpage}--\pageref{lastpage}}
\maketitle

\begin{abstract}
Excursion set theory is a powerful and widely used tool for describing the distribution of dark matter haloes, but it is normally applied with simplifying approximations. We use numerical sampling methods to study the mass functions predicted by the theory without approximations. With a spherical top-hat window and a constant $\delta=1.5$ threshold, the theory accurately predicts mass functions with the $M_{200}$ mass definition, both unconditional and conditional, in simulations of a range of matter-dominated cosmologies. For $\Lambda$CDM at the present epoch, predictions lie between the $M_\mathrm{200m}$ and $M_\mathrm{200c}$ mass functions. In contrast, with the same window function, a nonconstant threshold based on ellipsoidal collapse predicts uniformly too few haloes. This work indicates a new way to simply and accurately evaluate halo mass functions, clustering bias, and assembly histories for a range of cosmologies. We provide a fitting function that accurately represents the predictions of the theory for a wide range of parameters.
\end{abstract}


%%%%%%%%%%%%%%%%%%%%%%%%%%%%%%%%%%%%%%%%%%%%%%%%%%

%%%%%%%%%%%%%%%%% BODY OF PAPER %%%%%%%%%%%%%%%%%%

\section{Introduction}

The theory of excursion sets\footnote{Also known as extended Press-Schechter theory.} is a powerful tool for understanding dark matter haloes and how their distribution connects to the cosmological initial conditions. First used to predict halo mass functions \citep{1991ApJ...379..440B}, excursion sets are also employed to study other aspects of the halo distribution, such as their merger rates \citep{1993MNRAS.262..627L} and spatial clustering \citep{1989MNRAS.237.1127C,1996MNRAS.282..347M} \citep[for a review, see][]{2007IJMPD..16..763Z}. The basic idea of the excursion set approach is that a particle's density environment in the initial conditions directly predicts its halo membership at later times. If the density contrast field $\delta(\vec x)\equiv[\rho(\vec x)-\bar\rho]/\bar\rho$ averaged on the mass scale $M$ exceeds a preset threshold $\dc$, then the particle initially at $\vec x$ is part of a halo of at least mass $M$ at a preset later time. The masses $M$ satisfying this condition comprise the \emph{excursion set} associated with the threshold $\dc$ \citep{10.1214/aoap/1019737664}. The mass of the particle's host halo is taken to be the largest mass in the excursion set.

In particular, consider the density contrast field $\delta^{(M)}(\vec x,t)$ extrapolated to the time $t$ using linear-order cosmological perturbation theory and averaged on the mass scale $M$. At fixed position $\vec x$ and time $t$, $\delta^{(M)}$ can be regarded as executing a random walk in decreasing $M$, starting from $\delta^{(\infty)}=0$. Then the particle's host mass is the $M$ for which $\delta^{(M)}$ first crosses the threshold $\dc$. To make calculations of the first-crossing distribution analytically tractable, the random walk is ordinarily approximated to be Markovian, i.e., each step is uncorrelated with other steps. If the threshold $\dc$ is taken to be the spherical collapse threshold, $\dc= 1.686$, \citet{1991ApJ...379..440B} showed that this leads to the halo mass function of \citet{1974ApJ...187..425P}.
Approximations to the first-crossing distribution with correlated steps have also been studied \citep{1990MNRAS.243..133P,2010ApJ...711..907M,2012MNRAS.420.1429P,2012MNRAS.423L.102M,2013MNRAS.433.3428F,2014MNRAS.438.2683M,2014MNRAS.443.1601M,2018MNRAS.478.5296N}.

A common refinement of this approach is to employ a non-constant threshold $\dc$. Motivated by the \citet{1996ApJS..103....1B} model of ellipsoidal collapse, \citet{2001MNRAS.323....1S} considered a threshold that depends on the three-dimensional shape of the tidal deformation tensor $T_{ij}\equiv -\partial_i\partial_j\phi$, where $\phi$ is a solution to $\nabla^2\phi=-\delta$. Ellipsoidal collapse is delayed by tidal forces, so the threshold $\dc$ becomes higher if $T_{ij}$ is significantly aspherical. The random walk is now in the six independent components of $T_{ij}^{(M)}$, and \citet{2001ApJ...555...83C}, \citet{2002MNRAS.329...61S}, and \citet{2007MNRAS.377..234S} tested the first-crossing distributions that result therefrom. To simplify the problem, however, \citet{2001MNRAS.323....1S} exploited how the typical anisotropy of $T_{ij}^{(M)}$ depends on $M$ to approximate a mass-dependent ``moving'' threshold, $\dc(M)$. Since smaller $M$ is associated with more ellipsoidal tides, $\dc$ is taken to be a decreasing function of $M$. Compared to a constant threshold, halo mass functions resulting from the moving threshold of \citet{2001MNRAS.323....1S} yield a better match to the results of numerical simulations, assuming uncorrelated steps, although \citet{2009ApJ...696..636R} found that this may not hold when steps are correlated. Further refinements to the ellipsoidal collapse threshold have also been explored \citep{2010A&A...518A..38A,2014MNRAS.445.4110L,2014MNRAS.445.4124B}, while recent machine-learning-based studies have
questioned whether anisotropic information is relevant at all to halo mass predictions \citep{2018MNRAS.479.3405L,2019MNRAS.490..331L,2020arXiv201110577L}.

In this work, we relax the approximations by considering the full 6-dimensional $T_{ij}^{(M)}$ and accounting fully for correlations between steps, under several choices of averaging window function. We use direct numerical sampling, which can be done quite efficiently \citep[e.g.][]{2018MNRAS.478.5296N}, to generate halo mass functions for a range of cosmologies. We compare these mass functions both to the standard analytical predictions and to simulation results. Specifically, we consider both the scale-free and the concordance $\Lambda$CDM cosmological simulations of \citet{2015ApJ...799..108D}. We test both unconditional and conditional mass functions.

We find that for the standard choice of window function -- the spherical top-hat in real space -- the ellipsoidal collapse threshold predicts too few haloes of every mass. In contrast, a constant $\dc=1.5$ threshold yields halo mass functions that closely match those of the scale-free simulations with the $M_{200}$ mass definition.
The predicted conditional mass functions related to halo clustering bias and assembly history are also generally accurate.
For $\Lambda$CDM, the same threshold yields predictions that lie between $M_\mathrm{200m}$ and $M_\mathrm{200c}$ mass functions. Other common choices of window function cannot match the simulation results as closely with either constant or ellipsoidal-collapse-motivated thresholds. Mass functions predicted by excursion set theory with the top-hat window and a constant threshold are nearly independent of the linear power spectrum, when they are considered as a function of the rms density variance $\sigma$, but they can exhibit significant variations for extreme spectral indices or when there are features in the power spectrum.

This article is organized as follows.
Section~\ref{sec:sampling} describes our approach for numerically sampling the trajectories $\delta^{(M)}$ and $T_{ij}^{(M)}$.
In Sec.~\ref{sec:massfunctions}, we use those sampled trajectories to generate halo mass functions, and we compare the outcomes for different thresholds and window functions.
In Sec.~\ref{sec:simulation}, we compare the excursion set mass functions to those derived from cosmological simulations, considering both unconditional and conditional mass functions.
In Sec.~\ref{sec:universal}, we explore the degree to which excursion set mass functions adhere to a universal form when they are expressed in terms of $\sigma$, and we provide a fitting function.
We present conclusions in Sec.~\ref{sec:conclusion}.
Appendix~\ref{sec:convergence} tests how well the mass functions that we calculate are numerically converged, while Appendix~\ref{sec:conditionaluniversal} details the extent to which the predicted conditional mass functions stick to the same universal form.

\section{Numerical sampling of excursions}\label{sec:sampling}

This section describes how we numerically sample the trajectories of $\delta$ or $T_{ij}$ as a function of the averaging scale.

\subsection{Spherical collapse}\label{sec:sc}

\begin{figure*}
	\centering
	\includegraphics[width=\linewidth]{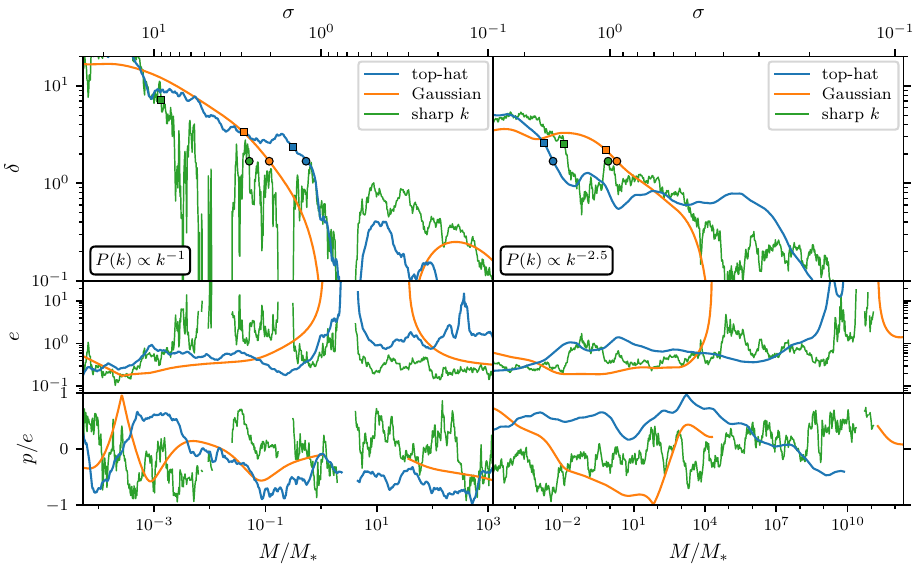}
	\caption{Example trajectories of the tidal tensor $T_{ij}$ as a function of averaging mass scale $M$ at a fixed position and time. We decompose $T_{ij}$ into the density contrast $\delta$ (upper panels), the ellipticity $e$ (middle panels), and the prolateness $p$ (lower panels, expressed in units of $e$). We consider a single trajectory for each of three averaging window functions (different colours) and two different scale-free power spectra (left-hand versus right-hand panels). The mass $M$ is expressed in units of the mass scale $M_*$ on which the rms variance in $\delta$ ($\sigma$, upper axis) is unity. Where $\delta<0$, we do not plot $e$ and $p$. The trajectories with the sharp $k$-space window (green) are extremely noisy, which reflects a lack of correlation between steps in $\delta$ as $M$ is varied. For the top-hat (blue) and Gaussian (orange) windows, the trajectories are much smoother. The circles mark first crossing of the spherical collapse threshold $\dc=1.686$ for each trajectory, while the squares mark first crossing of the ellipticity- and prolateness-dependent ellipsoidal collapse threshold $\dc=1.686\fec(e,p)$.}
	\label{fig:example_excursions}
\end{figure*}

For simplicity, we begin by restricting our consideration to the density contrast $\delta(\vx)$, which is a function of comoving position $\vx$. Let $\delta(\vk)=\int\diff^3\vx\, \e^{-\I \vk\cdot\vx}\delta(\vx)$ be its Fourier transform. The density contrast averaged on radius $r$ with the window function $W$ is
\begin{equation}
    \delta^{(r)}(\vx) = \int\frac{\diff^3\vk}{(2\pi)^3}\e^{\I \vk\cdot\vx}\delta(\vk)W(kr),
\end{equation}
whence it follows from the definition of the power spectrum $P(k)$ that
\begin{align}
    \left\langle\delta^{(r)}(\vx)\delta^{(r^\prime)}(\vx)\right\rangle 
    &= \int\frac{\diff^3\vk}{(2\pi)^3}P(k)W(kr)W(kr^\prime)
    \\\label{cov_dd}&= \int_0^\infty\frac{\diff k}{k}\mathcal{P}(k)W(kr)W(kr^\prime).
\end{align}
Here $\mathcal{P}(k)\equiv [k^3/(2\pi^2)]P(k)$ is the dimensionless power spectrum. In general, $W(x)$ is a function ranging from 1 for $x\ll 1$ to 0 for $x\gg 1$.

If we discretize the windowing radii $r$ into a sequence
$r_1<r_2<...<r_N$,
then $\langle\delta^{(r_a)}(\vx)\delta^{(r_b)}(\vx)\rangle$ is the $N\times N$ covariance matrix of the vector of $\delta^{(r_a)}\equiv\delta_a$. Under this discretization, a trajectory in $\delta$ is just a random vector $\delta_a$ distributed according to a $N$-variate Gaussian distribution with mean 0 and covariance $\langle \delta_a \delta_b\rangle$ given in accordance with Eq.~(\ref{cov_dd}). There are many ways to sample random $\delta_a$ from such a distribution \citep[e.g.][]{2018MNRAS.478.5296N}; we follow \citet{2019PhRvD.100b3523D} in diagonalizing
\begin{equation}\label{diagonalize}
    %\left\langle\delta^{(r_a)}(\vx)\delta^{(r_b)}(\vx)\right\rangle
    \langle\delta_a\delta_b\rangle = \sum_{c=1}^N A_{ac}\lambda_c A_{bc},
\end{equation}
where $A$ is an orthogonal matrix and $\lambda_a$ are the eigenvalues of $\langle\delta\delta\rangle$.
A randomly sampled trajectory is then given by
\begin{equation}\label{sample}
\delta_a=\sum_{b=1}^N A_{ab}w_b,    
\end{equation}
where each $w_b$ is sampled from the univariate Gaussian distribution of mean 0 and variance $\lambda_b$.

The upper panels of Fig.~\ref{fig:example_excursions} show examples of density trajectories sampled for three choices of spherical window function $W(x)$:
\begin{enumerate}
    \item the top-hat window function, $W(x)=(3/x^3)(\sin x-x\cos x)$, which cuts off sharply in real space;
    %For windowing radius $r$, the associated mass is $M=(4\pi/3)\rho_0 r^3$, where $\rho_0$ is the comoving matter density.
    \item the Gaussian window function, $W(x)=\e^{-x^2/2}$, which is smooth in both real and Fourier space; and
    %For a windowing radius $r$, the mass is $M=(2\pi)^{3/2}\rho_0 r^3$.
    \item the sharp $k$-space window function, $W(x)=1$ if $x<1$ and 0 otherwise.
    %For windowing radius $r$, we follow \citet{1993MNRAS.262..627L} in taking the mass to be $M=6\pi^2 \rho_0 r^3$.
\end{enumerate}
We express these trajectories in terms of the windowing mass scale $M\propto r^3$.
We consider two power spectra: $P(k)\propto k^n$ with $n=-1$ (left) and $n=-2.5$ (right). The windowing mass scales are expressed in units of $M_*$, the mass scale associated with the windowing radius $r_*$ such that $\sigma(r_*)=1$. Here
\begin{equation}\label{sigma}
    \sigma^2(r)=\int_0^\infty\frac{\diff k}{k}\mathcal{P}(k)W^2(kr)
\end{equation}
is the variance of $\delta^{(r)}(\vec x)$, the density field averaged on the radius $r$. Note that $M_*$ depends on the window function. When results are expressed in these units, the normalization of the power spectrum $P(k)$ is irrelevant, and the normalization of the windowing mass scale $M$ is also irrelevant.

Figure~\ref{fig:example_excursions} shows example trajectories in $\delta$, illustrating typical behaviour with the different window functions. For the sharp $k$-space window, steps in $\delta$ as $M$ is varied are uncorrelated, which leads to an extremely noisy trajectory. In contrast, the (real-space) top-hat window (blue) leads to much less noisy trajectories, and those with the Gaussian window (orange) are smooth.
It is interesting to note how the impact of Gaussian windowing is not intermediate between top-hat and sharp-$k$ windows, despite the mathematical sense in which the Gaussian window is intermediate between the other two.

\subsection{Ellipsoidal collapse}\label{sec:ec}

We now extend the above treatment to consider trajectories in the tidal tensor $T_{ij}$. In terms of the Fourier-transformed density contrast, $T_{ij}$ has the explicit expression
\begin{equation}
    T_{ij}(\vx)=\int\frac{\diff^3\vk}{(2\pi)^3}\e^{\I \vk\cdot\vx}\delta(\vk)\frac{k_i k_j}{k^2}.
\end{equation}
Since it is a symmetric $3\times 3$ matrix, it has six independent components. Three of them can be taken to be the eigenvalues $\lambda_1\geq\lambda_2\geq\lambda_3$, which describe the strength of tidal deformation along principal axes, while the remaining three degrees of freedom orient those axes. Note that $\delta=\lambda_1+\lambda_2+\lambda_3$. Following standard terminology for ellipsoidal collapse \citep[e.g.][]{1996ApJS..103....1B}, we define the ellipticity $e\equiv (\lambda_1-\lambda_3)/(2\delta)$ and prolateness $p\equiv (\lambda_1+\lambda_3-2\lambda_2)/(2\delta)$; thus $\delta$, $e$, and $p$ parametrize the eigenvalues of $T$.

The tidal tensor averaged on the radius $r$ with a window function $W$ is straightforwardly
\begin{equation}
    T_{ij}^{(r)}(\vx)=\int\frac{\diff^3\vk}{(2\pi)^3}\e^{\I \vk\cdot\vx}\delta(\vk)\frac{k_i k_j}{k^2}W(kr),
\end{equation}
from which it follows that
\begin{align}\label{Tcov0}
    \left\langle T_{ij}^{(r)}(\vx)T_{kl}^{(r^\prime)}(\vx)\right\rangle
    &= \int\frac{\diff^3\vk}{(2\pi)^3}P(k)\frac{k_i k_j k_k k_l}{k^4}W(kr)W(kr^\prime)
    \\\label{Tcov}
    &= \frac{\delta_{ij}\delta_{kl}\!+\!\delta_{ik}\delta_{jl}\!+\!\delta_{il}\delta_{jk}}{15}\left\langle\delta^{(r)}(\vx)\delta^{(r^\prime)}(\vx)\right\rangle, 
\end{align}
where $\delta_{ij}$ is the Kronecker delta (equal to 1 if $i=j$ and 0 otherwise). Equation~(\ref{Tcov}) results from carrying out the angular integrals in Eq.~(\ref{Tcov0}).
If we discretize the windowing radii $r_1<r_2<...<r_N$ again, then $\langle T_{ij}^{(r_a)}(\vx)T_{kl}^{(r_b)}(\vx)\rangle$ is a $6N\times 6N$ covariance matrix, since $T$ has 6 independent components. By diagonalizing this covariance matrix, we can sample random trajectories $\delta_a$, $e_a$, $p_a$ using the same methods as in Sec.~\ref{sec:sc}. The trajectories in Fig.~\ref{fig:example_excursions} were generated using this approach, and the lower panels show the ellipticity $e$ and prolateness $p$ as a function of the averaging mass scale $M$.

\subsection{Conditional trajectories}\label{sec:conditionalsampling}

We will also have occasion to sample trajectories in $\delta^{(r)}$ conditioned on it taking a particular value $\delta^{(\tilde r)}=\tilde\delta$ when windowed on a chosen scale $\tilde r$.
Given the discretization scheme $r_1<...<r_N$ again, the conditional distribution of the $\delta_a\equiv\delta^{(r_a)}$ is Gaussian with mean
\begin{equation}
    \left\langle \delta_a\right\rangle\Big|_{\delta^{(\tilde r)}=\tilde\delta} = \frac{\langle\delta^{(r_a)}\delta^{(\tilde r)}\rangle}{\sigma^2(\tilde r)}\tilde\delta
\end{equation}
and covariance
\begin{equation}
    \left\langle \Delta\delta_a \Delta\delta_b \right\rangle\Big|_{\delta^{(\tilde r)}=\tilde\delta} = \left\langle \delta_a \delta_b \right\rangle - \frac{\langle\delta^{(r_a)}\delta^{(\tilde r)}\rangle \langle\delta^{(r_b)}\delta^{(\tilde r)}\rangle}{\sigma^2(\tilde r)}
\end{equation}
(see Eqs. \ref{cov_dd} and~\ref{sigma}). Here $\Delta\delta_a \equiv \delta_a-\left\langle \delta_a\right\rangle|_{\delta^{(\tilde r)}=\tilde\delta}$. These results follow from a classic theorem on conditional Gaussian distributions \citep[e.g. Appendix~D of][]{1986ApJ...304...15B}. Similarly to Sec.~\ref{sec:sc}, we can diagonalize
\begin{equation}\label{diagonalize_cond}
    \left\langle \Delta\delta_a \Delta\delta_b \right\rangle\Big|_{\delta^{(\tilde r)}=\tilde\delta} = \sum_{c=1}^N A_{ac}\lambda_c A_{bc},
\end{equation}
where $A$ is an orthogonal matrix and $\lambda_a$ are the eigenvalues of $\langle\Delta\delta\Delta\delta\rangle|_{\delta^{(\tilde r)}=\tilde\delta}$.
A random trajectory is now given by
\begin{equation}\label{sample_cond}
\delta_a=\left\langle \delta_a\right\rangle\Big|_{\delta^{(\tilde r)}=\tilde\delta} + \sum_{b=1}^N A_{ab}w_b,
\end{equation}
where each $w_b$ is sampled from the Gaussian distribution of mean 0 and variance $\lambda_b$.

\section{Mass functions from excursion set theory}\label{sec:massfunctions}

In the excursion set approach, a particle is deemed to belong to a halo of mass $M$ if that is the largest mass scale for which the (linearly evolved) density contrast $\delta$ averaged around that particle's location exceeds some threshold $\dc$. That is, the halo mass is the location of first crossing of the threshold, if the trajectory in $\delta$ is viewed as a random walk in decreasing averaging scale $M$.
The circles in the upper panels of Fig.~\ref{fig:example_excursions} mark these first crossings if the spherical collapse threshold $\dc= 1.686$ is adopted.
We also consider the ellipsoidal collapse threshold $\dc=1.686\fec(e,p)$, where $\fec$ is the solution to
\begin{equation}\label{fec}
    \fec=1+0.47\left[5(e^2-p|p|)\fec^2\right]^{0.615},
\end{equation}
as approximated by \citet{2001MNRAS.323....1S}. Note that $\fec\geq 1$. The squares in Fig.~\ref{fig:example_excursions} mark each trajectory's first crossing of the ellipsoidal collapse threshold.
With respect to the horizontal axis, the circles and squares indicate the mass of the particle's host halo, as determined by spherical and ellipsoidal collapse, respectively.

We discretize the windowing radii $r_1<r_2<...<r_N$ such that $\sigma(r_N)=0.28$ (see Eq.~\ref{sigma}). This choice ensures that the probability is negligible that a trajectory would already exceed the collapse threshold $\dc$ at or above the maximum radius $r_N$, which is important because in that case the first crossing would be missed. We consider power spectra $P(k)\propto k^n$ with $n=-1$, $n=-1.5$, $n=-2$, and $n=-2.5$; the lower limit $r_1$ of the discretization is taken such that $\sigma(r_1)$ ranges from about 400 in the first case to about 30 in the last. When we sample one-dimensional trajectories in $\delta$ alone, we use $N=3200$ logarithmically spaced averaging radii, while to sample six-dimensional trajectories in $T_{ij}$, we reduce this number to $N=800$.\footnote{We discuss later (and in Appendix~\ref{sec:convergence}) the impact of the choice of $N$. Using \textsc{numpy} matrix operations \citep{harris2020array}, the author's personal computer samples about 400 trajectories in $\delta$ per second for $N=3200$ or about 150 trajectories in $T_{ij}$ per second for $N=800$ (including the decomposition into $\delta$, $e$, and $p$). See also \citet{2018MNRAS.478.5296N} for a potentially faster approach.} The spacing between successive windowing mass scales $M\propto r^3$ is listed in Table~\ref{tab:numerical} for each power spectrum.

\begin{table}
	\centering
	\caption{Spacing of steps in the top-hat window mass scale $M$ for the different power spectra. We also show the spacing in $\sigma$ (Eq.~\ref{sigma}). For other window functions, the steps are different by under 5 per cent. For one-dimensional trajectories in $\delta$ alone, we use much tighter spacing than for six-dimensional trajectories in $T_{ij}$ at similar computational expense.}
	\label{tab:numerical}
	\begin{tabular}{lcccc}
		\hline
        & \multicolumn{2}{c}{$\delta$ (spherical)} & \multicolumn{2}{c}{$T_{ij}$ (ellipsoidal)} \\
        $P(k)$ & $\Delta\ln M$ & $\Delta\ln\sigma$ & $\Delta\ln M$ & $\Delta\ln\sigma$ \\
		\hline
$\propto k^{-1}$ & 0.0056 & 0.0019 & 0.0224 & 0.0075 \\
$\propto k^{-1.5}$ & 0.0065 & 0.0016 & 0.0259 & 0.0065 \\
$\propto k^{-2}$ & 0.0079 & 0.0013 & 0.0317 & 0.0053 \\
$\propto k^{-2.5}$ & 0.0127 & 0.0011 & 0.0510 & 0.0042 \\
		\hline
	\end{tabular}
\end{table}

\begin{figure*}
	\centering
	\includegraphics[width=\linewidth]{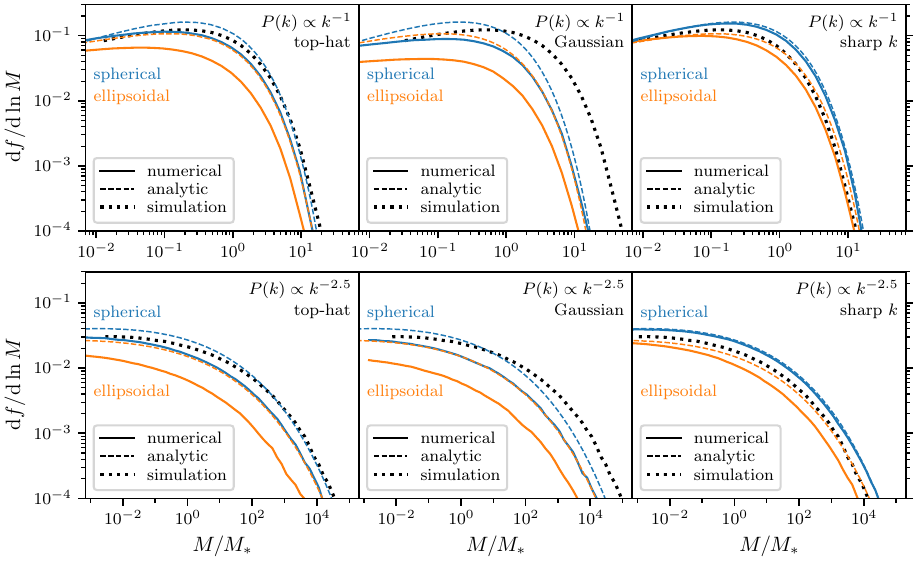}
	\caption{
 Differential fraction of particles $\diff f/\diff\ln M$ in haloes of mass $M$, as evaluated via excursion set theory through our numerical procedure (solid curves). We consider both the constant spherical-collapse threshold $\dc=1.686$ (blue curves) and the ellipsoidal collapse threshold $\dc=1.686\fec(e,p)$ (orange curves; see Eq.~\ref{fec}).
 The upper panels are for a $P(k)\propto k^{-1}$ power spectrum, while the lower panels are for $P(k)\propto k^{-2.5}$. From left to right, we show results for the top-hat, Gaussian, and sharp $k$-space window functions. The $\diff f/\diff\ln M$ mass functions are expressed in units of the characteristic mass scale $M_*$ on which the rms variance is 1; note that $M_*$ depends on the window function.
 For comparison, the dashed curves show the classic analytic predictions \citep[][]{1974ApJ...187..425P,2001MNRAS.323....1S}, which are the same in the left-hand, centre, and right-hand panels. Meanwhile, the black dotted curves show the $M_{200}$ mass functions from the simulations of \citet{2015ApJ...799..108D} (which depend on the window only because $M_*$ does).
 The numerically evaluated mass functions with top-hat and Gaussian windows differ significantly from the analytic approximations. For these windows, the ellipsoidal collapse threshold yields too few haloes of every mass compared to the simulation results.
 }
	\label{fig:hmf_demo}
\end{figure*}

For each power spectrum, we sample $10^5$ trajectories in $\delta$ alone and $10^5$ trajectories in $T_{ij}$. For each trajectory, we find the largest windowing scale for which $\delta$ exceeds the threshold for spherical or ellipsoidal collapse, and then we attempt to interpolate the precise scale at which the crossing occurred.
The first crossing sets the mass $M$ of the halo hosting the particle associated with the trajectory in question. Since we are sampling trajectories associated with arbitrary points in the initial density field, our sample is of arbitrary dark matter particles. Therefore, the distribution of first-crossing masses is precisely the differential fraction $\diff f/\diff M$ of all mass that resides in haloes of mass $M$. We will generally present halo mass functions as $\diff f/\diff\ln M$, the differential mass fraction per logarithmic interval in halo mass, but note that the more commonly discussed differential halo number density is related by
\begin{equation}\label{dndlogM}
    \frac{\diff n}{\diff\ln M}=\frac{\bar\rho}{M}\frac{\diff f}{\diff\ln M}.
\end{equation}

A great convenience of considering scale-free cosmologies, with $P(k)\propto k^n$, is that a change in time is equivalent to a change in mass scale.\footnote{Other studies that took advantage of self-similarity in scale-free cosmologies include \citet{1988MNRAS.235..715E,1994MNRAS.271..676L,1999ApJ...517L...5L,2008MNRAS.385..545K,2009MNRAS.395.1950E,2015ApJ...799..108D,2017MNRAS.465L..84L,2019ApJ...871..168D,2020ApJ...903...87D,2021MNRAS.501.5051J,2021MNRAS.501.5064L}.} This means that we can improve the statistical precision of this calculation -- particularly at the large-mass end -- by stacking the distributions of first-crossing masses evaluated at different times. We assume scale-independent growth (valid for a dark matter-dominated universe) and adopt growth factors $D$ ranging from $\sigma(r_N)/\sigma(r_1)$ to 1 separated by factors of 1.03, where $\sigma(r_1)\sim \mathcal{O}(100)$ and $\sigma(r_N)=0.28$ are the rms variance at the minimum and maximum window scales, respectively, as described above. Thus, at the smallest $D$ (earliest time), no haloes are expected within the resolution limit implied by the choice of window scales. By uniformly scaling each previously sampled trajectory in $\delta$ by the growth factor $D$ (or equivalently scaling the thresholds by $1/D$), we obtain the first-crossing mass distribution at the time when the growth factor was $D$. The characteristic mass scale $M_*(D)$ at this earlier time is defined to be the mass associated with the window radius $r_*$ such that $D\sigma(r_*)=1$ (see Eq.~\ref{sigma}). In units of $M_*(D)$, the underlying first-crossing mass distributions at different $D$ must be all exactly the same.

We count first-crossing masses $M/M_*(D)$ in logarithmic bins of width $\Delta\ln M=0.33$. For each mass bin, we stack the counts from all growth factors $D$ for which the bin lies fully between the lower and upper mass limits, $M_1/M_*(D)$ and $M_N/M_*(D)$, where $M_i$ is the mass associated with the window radius $r_i$.
Since $M_*(D)$ grows with $D$, the distributions at lower $D$ (earlier times) tend to improve the count of first-crossing masses at high $M/M_*(D)$.
We will show in Sec.~\ref{sec:simulation} that the statistical uncertainty in $\diff f/\diff\ln M$ resulting from this procedure is mostly at the per cent level. Note that the uncertainty of the count in each mass bin is not simply Poissonian, because the counts contributed by different $D$ are correlated.

The solid curves in Fig.~\ref{fig:hmf_demo} show the $\diff f/\diff\ln M$ that result from this calculation, for both the spherical collapse (blue) and ellipsoidal collapse (orange) thresholds. We consider three different window functions $W$ (different columns) and the two power spectra $P(k)\propto k^n$ with $n=-1$ and $n=-2.5$ (different rows). For comparison, we also show as dashed lines the standard analytic predictions,
\begin{equation}\label{sc}
    \left.\frac{\diff f}{\diff\ln M}\right|_\text{sc}
    =
    \sqrt{\frac{2}{\pi}}\frac{\dc}{\sigma_M}\exp\left(-\frac{\dc^2}{2\sigma_M^2}\right)
    \left|\frac{\diff\ln\sigma_M}{\diff\ln M}\right|
\end{equation}
\citep{1974ApJ...187..425P,1991ApJ...379..440B} for the spherical collapse threshold and approximately
\begin{equation}\label{ec}
    \left.\frac{\diff f}{\diff\ln M}\right|_\text{ec}=0.3222\left[1+\left(\frac{\dc}{\sigma_M}\right)^{-0.6}\right]
    \left.\frac{\diff f}{\diff\ln M}\right|_\text{sc}
\end{equation}
\citep{2001MNRAS.323....1S} for the ellipsoidal collapse threshold. Here $\sigma_M$ is shorthand for $\sigma(r)$ (Eq.~\ref{sigma}) evaluated for the radius $r\propto M^{1/3}$ associated with the mass scale $M$.
For spherical collapse with the sharp $k$-space window, our numerical calculation matches the analytic prediction almost exactly. This outcome is expected since the analytic prediction is exact under the assumption that steps in $\delta$ are uncorrelated, and that assumption is satisfied for sharp $k$-space windowing. For ellipsoidal collapse with the same window, the analytic prediction is close to our numerical result but does not exactly match it; this outcome reflects that Eq.~(\ref{ec}) is only approximate even under the assumption that steps are uncorrelated.

The top-hat and Gaussian windows yield more remarkable outcomes. For the top-hat window, the numerically evaluated mass function from spherical collapse nearly matches the analytic prediction \emph{for ellipsoidal collapse} (not spherical collapse).
A similar outcome was noticed before by \citet{2014MNRAS.443.1601M}.
Meanwhile, the numerically evaluated mass function from ellipsoidal collapse is completely different, predicting significantly too little mass in haloes of every mass \citep[as noted by][]{2009ApJ...696..636R}. Similar behaviour is also true for the Gaussian window. To the extent that ellipsoidal collapse improves the match between excursion set mass functions and simulation results \citep{2001MNRAS.323....1S}, this outcome suggests that the improvement may only have been an artifact of the assumption that steps in the trajectories of $\delta$ are uncorrelated. In the next section, we will test the degree to which excursion set theory with a constant threshold $\dc$ can predict mass functions that match simulation results.

There is one important source of error in our numerically evaluated mass functions. Due to the discretization of window scales, it is possible for the procedure to miss a sufficiently ``brief'' crossing of the $\delta=\dc$ threshold. That is, we may have $\delta^{(r_i)}<\dc$ and $\delta^{(r_{i+1})}<\dc$, but $\delta^{(r)}>\dc$ for some $r_i<r<r_{i+1}$. This effect tends to improperly shift the first-crossing distribution to smaller window scales, an outcome that can be seen in the comparison between the analytic prediction for spherical collapse and the numerical counterpart in the rightmost panels of Fig.~\ref{fig:hmf_demo}. The numerical distribution (solid blue curve) lies slightly below the analytic one (dashed blue curve) for most of the mass range, except for the lowest masses for the $P(k)\propto k^{-1}$ power spectrum, where the numerical curve is slightly above the analytic one. In Appendix~\ref{sec:convergence}, we test the degree to which our numerically evaluated mass functions are converged with respect to the resolution of the discretization scheme. We find that the mass functions converge much more readily for the top-hat and Gaussian windows than for the sharp-$k$ window. 
This result can be understood by appealing to the example trajectories in Fig.~\ref{fig:example_excursions}. Trajectories with the sharp-$k$ window are extremely noisy, which makes ``brief'' threshold crossings likely. In contrast, trajectories with the top-hat and Gaussian windows are much smoother. Although our numerically evaluated mass functions with the sharp $k$-space window may be inaccurate at the 10 per cent level, there is no reason to expect the top-hat and Gaussian counterparts to be inaccurate to any comparable degree.

\section{Comparison with simulations}\label{sec:simulation}

\subsection{Scale-free cosmology}\label{sec:selfsim}

The dotted curves in Fig.~\ref{fig:hmf_demo} compare our results to mass functions derived from cosmological simulations. We use the publicly available halo catalogues for the scale-free simulations of \citet{2015ApJ...799..108D}, which are part of the Erebos simulation suite \citep{2020ApJS..251...17D}.
There are four simulations, initialized with power spectra $P(k)\propto k^n$ with $n=-1$, $-1.5$, $-2$, and $-2.5$, although we only represent two of them in Fig.~\ref{fig:hmf_demo}.
These simulations involve dark matter particles only and adopt flat, matter-dominated (Einstein-de Sitter) cosmologies.
Catalogued haloes are identified with the \textsc{rockstar} halo finder \citep{2013ApJ...762..109B}, and the calculation of their masses is described by \citet{2020ApJS..251...17D}. We use the $M_{200}$ mass definition, which is the mass enclosed within a sphere centred on the halo's inner cusp, where that sphere is the largest one that encloses average density 200 times the cosmological average. We consider only field haloes (as opposed to subhaloes).\footnote{Halo mass functions from the same simulations were previously analysed by \citet{2020ApJ...903...87D}. While scale-free simulations are less widely studied, mass functions from $\Lambda$CDM simulations in the same \citet{2020ApJS..251...17D} suite were shown by \citet{2021MNRAS.500.3309M} to agree with a wide range of independent simulations.}
Note that the simulated mass functions in Fig.~\ref{fig:hmf_demo} depend on the window function only because they are expressed in units of the characteristic mass scale $M_*$ (on which the rms variance of $\delta$ is 1), which depends on the window function.

For each simulation, a range of snapshots are available, separated by factors of about 1.03 in the scale factor $a$ and hence in the growth factor $D$. The snapshots span a total factor in $a$ ranging from 6 to 11, depending on the simulation. Since these simulations involve scale-free cosmologies, different snapshots sample different portions of the same halo mass distribution, as we discussed in the previous section. The mass scale $M_*$ grows over time, so earlier snapshots sample haloes of larger $M/M_*$ while later snapshots sample haloes of smaller $M/M_*$.
The simulations involved $1024^3$ particles in a periodic box, and Table~\ref{tab:simulation} lists how the particle mass and the total box mass compare to $M_*$ for the earliest and latest snapshots of each simulation.
We count haloes in the same bins of width $\Delta\ln M=0.33$ as in the previous section. Since haloes with too few particles can be influenced by discreteness or other resolution artifacts, we follow \citet{2020ApJ...903...87D} in only considering haloes for which $M_{200}$ is more than 500 times the particle mass.

Figure~\ref{fig:hmf_demo} shows that with the top-hat window (left-hand panels), the excursion set mass functions with the spherical collapse threshold $\dc=1.686$ (solid blue curve) lie close to the simulation mass functions but are offset in mass. However, there is no reason to adopt precisely the spherical collapse threshold, which is associated with the time at which a spherical shell collapses to radius 0 (under simplifying assumptions). We expect in general that a particle will cross a halo's $R_{200}$ boundary (the radius of the 200-times-overdense sphere), to contribute to the halo's $M_{200}$, significantly earlier.
Also, if the true threshold $\dc$ exhibits stochasticity, this can be statistically equivalent to a lower threshold without stochasticity \citep{2010ApJ...717..515M}.
Thus, it is natural to choose a lower threshold, which would shift the excursion set mass functions to the right (higher masses).

\begin{table}
	\centering
	\caption{How the mass resolution limits for the scale-free simulations of \citet{2015ApJ...799..108D} employed in this work compare to the characteristic mass scale $M_*$ (on which the rms variance of $\delta$ is 1). Here we evaluate $M_*$ using a top-hat window.
 }
	\label{tab:simulation}
	\begin{tabular}{lcccc}
		\hline
        & \multicolumn{2}{c}{earliest snapshot} & \multicolumn{2}{c}{latest snapshot} \\
        $P(k)$ & $\frac{\text{box mass}}{M_*}$ & $\frac{\text{particle mass}}{M_*}$ & $\frac{\text{box mass}}{M_*}$ & $\frac{\text{particle mass}}{M_*}$ \\
		\hline
$\propto k^{-1}$ & $7.7 \times 10^{6}$ & $7.2 \times 10^{-3}$ & $2.2 \times 10^{4}$ & $2.1 \times 10^{-5}$ \\
$\propto k^{-1.5}$ & $2.0 \times 10^{8}$ & $1.9 \times 10^{-1}$ & $1.5 \times 10^{4}$ & $1.4 \times 10^{-5}$ \\
$\propto k^{-2}$ & $3.0 \times 10^{10}$ & $2.8 \times 10^{1}$ & $1.6 \times 10^{4}$ & $1.5 \times 10^{-5}$ \\
$\propto k^{-2.5}$ & $2.4 \times 10^{13}$ & $2.2 \times 10^{4}$ & $5.0 \times 10^{3}$ & $4.7 \times 10^{-6}$ \\
	\end{tabular}
\end{table}

\begin{figure*}
	\centering
	\includegraphics[width=\linewidth]{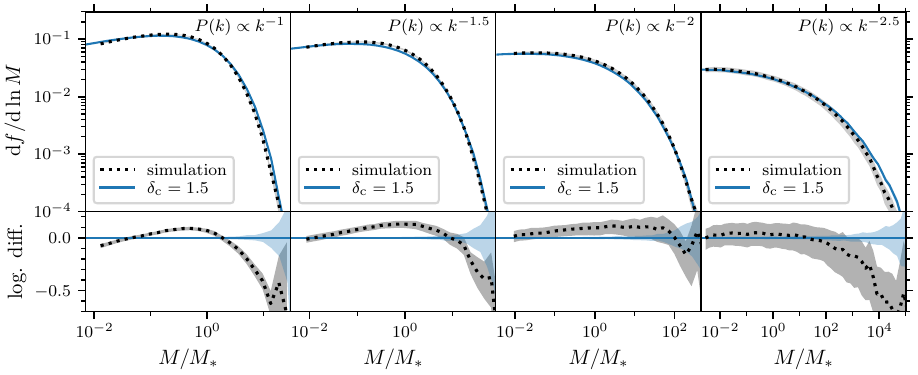}
	\caption{Comparing halo mass functions from excursion set theory with simulation results. As in Fig.~\ref{fig:hmf_demo}, the upper panels show the differential fraction $\diff f/\diff\ln M$ of mass that resides in haloes of mass $M$. For the excursion set predictions, we use top-hat windowing and adopt the threshold $\dc=1.5$. For the simulations, we adopt the $M_{200}$ mass definition. Different panels represent different scale-free power spectra.
	Shading indicates the 90 per cent confidence uncertainty bands. The lower panels show the (base $\e$) logarithmic differences $\Delta\ln(\diff f/\diff\ln M)$ between simulations and predictions.
	Excursion set theory with $\dc=1.5$ appears to predict the simulation mass functions in all cases to a high level of accuracy.}
	\label{fig:hmf_tophat}
\end{figure*}

Figure~\ref{fig:hmf_tophat} shows excursion set mass functions with the top-hat window and a lower threshold $\dc=1.5$. Here we consider all of the four different power spectra $P(k)\propto k^n$ with $n=-1$, $-1.5$, $-2$, and $-2.5$ (different panels). In each case, we compare to the mass function from the respective simulation.
We include 90 per cent confidence uncertainty bands, which we estimate by bootstrapping. For the excursion set predictions, we resample $10^5$ samples (with replacement) from the previously sampled trajectories. For the simulations, we split the simulation volume into 64 cubes and resample 64 at random (with replacement). In both cases, we evaluate $\diff f/\diff\ln M$ for 101 such resamplings, and the uncertainty bands extend between the 5th and 95th percentiles. Where $\diff f/\diff\ln M$ is close to its maximum value, the uncertainty in the excursion set predictions is around 1 per cent, too small to be visible.

The match between simulations and predictions is quite close, remaining at the 10 per cent level except where $\diff f/\diff\ln M$ drops off steeply at high masses.
Moreover, we will show in Sec.~\ref{sec:conditional} that the discrepancy at high masses for $P(k)\propto k^{-2.5}$ is an artificial consequence of the finite simulation box size.
The success of this model is particularly notable because it has only a single parameter, the threshold $\dc$.
We emphasize that once $\dc$ is chosen, there is no further freedom, and that the same value, $\dc=1.5$, works well for all four of the power spectra.
This constant threshold evidently suffices for excursion set theory to predict halo mass functions accurately, and no correction for ellipsoidal collapse appears to be necessary.

The possibility that ellipsoidal collapse does not improve excursion set predictions of halo masses may come as a surprise.
The ellipsoidal collapse threshold in Eq.~(\ref{fec}) is known to accurately predict the outcome of the collapse of a local maximum in the \emph{unwindowed} density field \citep{2019PhRvD.100b3523D,2023MNRAS.518.3509D,2023arXiv230905707O}.
Ellipsoidal collapse may also be relevant to the association of haloes with local maxima in the windowed initial density field \citep[e.g.][]{1996ApJS..103...41B,2016arXiv161103619C}.
However, the standard excursion set theory considered in this work is conceptually quite different. Rather than tracking haloes, it tags individual particles with the mass of their host halo. The threshold $\dc$ is associated with the time at which a particle becomes part of a halo of this mass, which could involve the particle's current halo growing through accretion but could also involve that halo accreting onto a larger host.
It is unclear the extent to which the ellipsoidal collapse threshold in Eq.~(\ref{fec}) -- which marks the time for all three axes of a homogeneous ellipsoid to collapse \citep[under simplifying assumptions about the behaviour of the first two axes as they approach collapse; see][]{1996ApJS..103....1B,2001MNRAS.323....1S} -- should be expected to apply to this problem.
Moreover, our result is consistent with the findings of \citet{2018MNRAS.479.3405L,2019MNRAS.490..331L,2020arXiv201110577L}, which showed that tidal shear and three-dimensional shape information do not improve machine-learning-based halo mass predictions compared to using spherically averaged density information alone.

The accuracy of the constant-threshold excursion set theory is also surprising because it cannot hold at the level of individual particles \citep[as pointed out by][]{1991ApJ...379..440B}. For example, neighboring particles are associated in general with haloes of different masses, even if those masses are large enough to imply that the particles belong to the same halo. This deficiency has motivated alternative approaches that attempt to identify the precise set of particles in the initial conditions that will belong to a halo at late times, often by associating haloes with local maxima in the windowed density field
\citep[e.g.][]{1990MNRAS.245..522A,1996ApJS..103....1B,1996ApJS..103...41B,1998ApJ...499..548M,2001MNRAS.327..721H,2011MNRAS.413.1961L,2012MNRAS.426.2789P,2012MNRAS.421..296R,2013MNRAS.431.1503P,2013MNRAS.430.1486R,2014MNRAS.438..878H,2016arXiv161103619C,2021MNRAS.508.3634M,2023MNRAS.523L...4M}. Nevertheless, the conceptual simplicity of the excursion set approach is a major advantage. Moreover, we will show next that the theory can accurately predict not only unconditional but also conditional mass functions, making it suitable for a wide range of applications.

\subsection{Conditional mass functions}\label{sec:conditional}

A powerful feature of excursion set theory is its capacity to predict halo mass functions in regions that have a specified density contrast $\delta$ at a larger mass scale. These conditional mass functions employ constrained trajectories in $\delta$, which can be sampled as described in Sec.~\ref{sec:conditionalsampling}. Conditional mass functions naturally enable treatments of halo clustering bias \citep[e.g.][]{1989MNRAS.237.1127C,1996MNRAS.282..347M,1999MNRAS.308..119S,2011MNRAS.411.2644M,2012MNRAS.419..132P,2014ApJ...782...44Z,2023arXiv231016093Z}. A conditional mass function can also be interpreted as the mass function of a given halo's progenitors at some earlier time \citep[e.g.][]{1993MNRAS.262..627L,2000MNRAS.319..168C,2007MNRAS.376..977G,2008MNRAS.383..557P,2010MNRAS.401.1796A,2013MNRAS.428.1774B,2014MNRAS.440..193J,2023MNRAS.521.3201N}. We now test the degree to which conditional mass functions predicted by the $\dc=1.5$ excursion set theory match the results of the scale-free simulations.

As periodic simulations of flat cosmologies, the simulations that we use enforce $\delta=0$ on the scale of the box.
Ordinarily, this constraint does not significantly influence structures at scales much smaller than the box \citep[e.g.][]{2006MNRAS.370..691P}, because the amplitudes of density contrasts are typically much larger at those scales than they would be expected to be at the box scale.
However, for $P(k)\propto k^{-2.5}$, the rms density contrast $\sigma$ scales as $\sigma\propto M^{-1/12}$. Since the smallest halo mass that we consider (500 simulation particle masses) is about $5\times 10^{-7}$ times the mass of the full simulation box, this means that between the smallest scale and the box scale of the $P(k)\propto k^{-2.5}$ simulation, there is only a factor of about 3 in $\sigma$.

\begin{figure}
	\centering
	\includegraphics[width=\columnwidth]{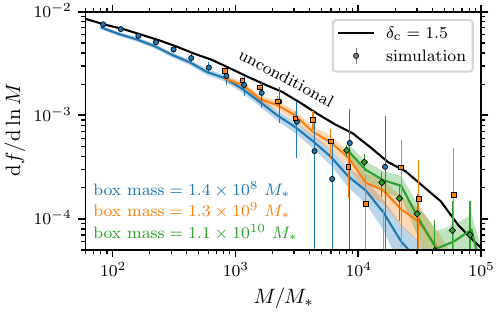}
	\caption{Testing how accurately excursion set theory with top-hat window and $\dc=1.5$ accounts for bias due to the simulation box size. The points (with $2\sigma$ Poisson uncertainty bars) are mass functions $\diff f/\diff\ln\sigma$ for the $P(k)\propto k^{-2.5}$ simulation at three different snapshots (different colours), specified by how the box size compares to the characteristic mass scale $M_*$ (at which $\sigma=1$). For visual convenience, points corresponding to the same mass bin have slightly different horizontal offsets for different simulations. The solid curves (with $2\sigma$ Poisson uncertainty bands) show the conditional mass functions predicted by excursion set theory for the same box size. For comparison, the black curve is the unconditional mass function.}
	\label{fig:bias}
\end{figure}

For the $P(k)\propto k^{-2.5}$ simulation, we consider three snapshots corresponding to box masses ranging from about $10^8$ to about $10^{10}$~$M_*$.
Using a set of window scales separated by $\Delta\ln\sigma=0.002$, we generate $10^6$ trajectories for each box size, conditioned on $\delta=0$ at the box scale.\footnote{One subtlety is that the simulation enforces $\delta=0$ in a cube and not a sphere. Since we consider only spherical windows, we approximate that $\delta=0$ in a sphere of radius equal to the box size.}
We count first-crossing masses (with $\dc=1.5$) in the same mass bins as in Sec.~\ref{sec:selfsim}. In Fig.~\ref{fig:bias}, we show both these conditional mass functions and the mass functions from the corresponding simulation snapshots. For comparison, we also repeat the unconditional mass function. The conditional mass functions appear to match the simulation results reasonably well. Evidently, the $\dc=1.5$ model accurately accounts for halo clustering bias, at least in this case.

\begin{figure}
	\centering
	\includegraphics[width=\columnwidth]{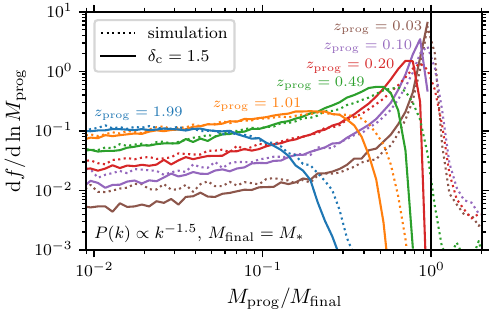}
	\caption{Testing how accurately excursion set theory with top-hat window and $\dc=1.5$ predicts progenitor mass functions. For field haloes of mass $M_\mathrm{final}=M_*$ at a scale factor $a$ in the $P(k)\propto k^{-1.5}$ simulation, the dotted curves show the differential fraction $\diff f/\diff\ln M_\mathrm{prog}$ of their mass that is determined to have been in field haloes of mass $M_\mathrm{prog}$ at the earlier scale factor $a/(1+z_\mathrm{prog})$. The solid curves show the corresponding predictions of the excursion set theory. Different colours correspond to different $z_\mathrm{prog}$. Predictions generally match the simulation results well. The principal discrepancy is at the largest masses, which may be related to backsplash haloes, as discussed in the text.}
	\label{fig:conditional}
\end{figure}

Next, we test whether the $\dc=1.5$ model can predict halo progenitor mass functions. For haloes of mass $M_\mathrm{final}=M_*$ at the scale factor $a$, the dotted curves in Fig.~\ref{fig:conditional} show the differential fraction of their mass that was in haloes of mass $M_\mathrm{prog}$ at the earlier scale factor $a/(1+z_\mathrm{prog})$. Specifically, we employ the $P(k)\propto k^{-1.5}$ simulation and consider field haloes between the masses $M_*/1.03$ and $1.03M_*$ at each scale factor $a$. Then we track the progenitors of those haloes and all of their subhaloes \citep[as determined by the \textsc{rockstar} and \textsc{consistent-trees} codes;][]{2013ApJ...762..109B,2013ApJ...763...18B} back to the scale factor $a/(1+z_\mathrm{prog})$ and determine the mass function of the host haloes of all of those progenitors (where a field halo is regarded as its own host). We count these progenitors in bins of width $\Delta\ln M_\mathrm{prog}=0.1$ and stack the counts from all scale factors $a$ for which there is a simulation snapshot.

\begin{figure*}
	\centering
	\includegraphics[width=\linewidth]{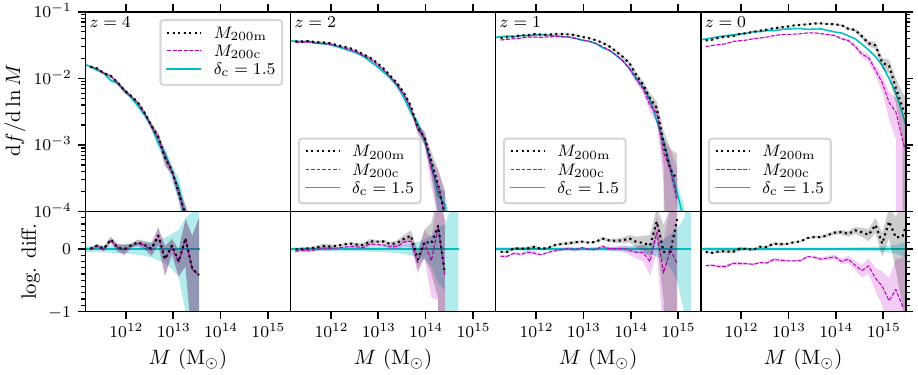}
	\caption{Testing excursion set theory predictions in a concordance cosmology. The upper panels show the differential fraction $\diff f/\diff\ln M$ of mass that resides in haloes of mass $M$. For the excursion set predictions (solid curves), we use top-hat windowing and the threshold $\dc=1.5$, as in Fig.~\ref{fig:hmf_tophat}. For the simulations (dotted curves), we consider the $M_\mathrm{200m}$ (black) and $M_\mathrm{200c}$ (magenta) mass definitions. Shading marks the estimated $2\sigma$ Poisson uncertainty for each curve. Different panels correspond to different redshifts. The lower panels show the (base $\e$) logarithmic differences $\Delta\ln(\diff f/\diff\ln M)$ between simulations and predictions. Excursion set theory with $\dc=1.5$ accurately predicts the simulation mass functions at high redshift, when matter dominates, but at low redshifts the prediction lies between the $M_\mathrm{200m}$ and $M_\mathrm{200c}$ mass functions.}
	\label{fig:hmf_concordance}
\end{figure*}

Excursion set theory predicts progenitor mass functions in the following way. For trajectories that first cross $\dc=1.5$ at the $M_\mathrm{final}$ window scale, we seek the distribution of first crossings of the higher threshold $(1+z_\mathrm{prog})\dc$. With a succession of window scales spaced by $\Delta\ln\sigma = 0.002$, we use the method of Sec.~\ref{sec:conditionalsampling} to sample trajectories that cross $\dc=1.5$ at the mass $M_\mathrm{final}=M_*$. To ensure that this crossing is the first, we include window scales above $M_\mathrm{final}$ up to the point that $\delta=\dc$ would represent a $5\sigma$ upward deviation, and then we reject any trajectories that exceed $\dc$ at any mass scale larger than $M_\mathrm{final}$. We obtain $10^5$ trajectories in this way. For each $z_\mathrm{prog}$, we find the first-crossing distribution of these trajectories for the threshold $(1+z_\mathrm{prog})\dc$, counting first crossings in the same bins of width $\Delta\ln M_\mathrm{prog}=0.1$ as we used for simulated haloes. The resulting conditional mass functions are shown in Fig.~\ref{fig:conditional} as solid curves.

The simulated progenitor mass functions in Fig.~\ref{fig:conditional} match the excursion set predictions reasonably well. The main difference is that the simulated mass functions tend to extend to slightly higher masses (including, for low $z_\mathrm{prog}$, $M_\mathrm{prog}>M_\mathrm{final}$). This outcome is likely connected to the existence of ``backsplash'' haloes that pass through a host halo before becoming (at least briefly) field haloes again \citep[e.g.][]{2021ApJ...909..112D}. This possibility means that the mass of a dark matter particle's field-halo host can decrease in time, which would explain why $M_\mathrm{prog}$ can exceed $M_\mathrm{final}$. However, in excursion set theory, the mass of a particle's field-halo host can only increase over time. A modified halo mass definition that accounts for backsplash haloes (either treating them as subhaloes of their previous hosts or taking them to be field haloes at all times until their last infall) may be needed to match excursion set predictions more precisely.

\subsection{Concordance cosmology}\label{sec:concordance}

Excursion set theory with the top-hat window and $\dc=1.5$ evidently predicts $M_{200}$ halo mass functions accurately for a range of scale-free cosmologies.
Figure~\ref{fig:hmf_concordance} now tests the model's predictions for a concordance ($\Lambda$CDM) cosmology. We compare the simulations of \citet{2015ApJ...799..108D} that were carried out with \citet{2014A&A...571A..16P} cosmological parameters, which are also part of the Erebos simulation suite \citep{2020ApJS..251...17D}. There are three $1024^3$-particle simulations with periodic box sizes 187, 373, and 746~Mpc. Since concordance cosmology includes dark energy, it is necessary to clarify the halo mass definition further. We consider both the $M_\mathrm{200m}$ mass definition, that of the sphere that has 200 times the average matter density, and the $M_\mathrm{200c}$ mass definition, that of the sphere of density 200 times the total (or critical) energy density. We consider four different redshifts (different panels), in each case stacking halo counts from all three simulations. As before, we count haloes in mass bins of width $\Delta\ln M=0.33$ and consider only haloes larger than 500 times the simulation particle mass.

To generate the excursion set predictions, we sample $4\times 10^5$ trajectories in $\delta$ between the masses $1.5\times 10^{10}$~M$_\odot$ (at which $\sigma\simeq 4$) and $1.5\times 10^{16}$~M$_\odot$ (at which $\sigma=0.28$). We take an interval of $\Delta\ln\sigma=0.002$, resulting in $N=1327$ window scales. We obtain first-crossing distributions at different redshifts $z$ by employing the growth factor
\begin{equation}
    D(z) = (1+z)^{-1}\,_2F_1(1/3,1;11/6;-\rhoM/\rhoL)
\end{equation}
\citep[e.g.][]{2011JCAP...10..010B}, where $_2F_1$ is the hypergeometric function and $\rhoM/\rhoL$ is the ratio of matter to dark energy density as a function of $z$. We normalize the top-hat window mass as $M=(4\pi/3)\rho_\mathrm{DM}r^3$, where $\rho_\mathrm{DM}$ is the comoving dark matter density, and we count the resulting halo masses in the same mass bins as we used for the simulated haloes.

Figure~\ref{fig:hmf_concordance} shows that the excursion set predictions closely match the simulation results at high redshifts (left-hand panels), as we expect from the results of the previous subsection, since matter dominates at these redshifts. At lower redshifts (right-hand panels), dark energy begins to dominate, leading the simulated mass functions with the two different mass definitions to diverge from each other. We find that the excursion set prediction does not match either simulated mass function but instead lands between them, although it tends to match the $M_\mathrm{200m}$ mass function at the lowest and highest masses. The influence of the mass definition on halo mass functions has been widely discussed \citep[e.g.][]{2001MNRAS.323....1S,2008ApJ...688..709T,2016MNRAS.456.2486D,2020ApJ...903...87D}, and it is possible that a different mass definition would yield a closer match at low redshifts to the excursion set theory predictions.\footnote{For example, we tested the time-dependent ``virial'' spherical-overdensity mass definition \citep[motivated by spherical collapse arguments, e.g.][]{1998ApJ...495...80B}. While it matches the $\dc=1.5$ prediction at $z=0$ for a large portion of the mass range, the match is not strong at all masses and redshifts.} It may also be appropriate to vary the threshold $\dc$ with redshift, due to the increasing influence of dark energy \citep[e.g.][]{1993MNRAS.262..627L}.

\section{Universality of mass functions?}\label{sec:universal}

When the halo mass function is expressed in units of $\sigma$ instead of $M$, excursion set theory with uncorrelated steps predicts that it is universal. That is, let $\diff f/\diff\ln\sigma$ be the differential mass fraction in haloes of the mass scale for which the rms density variance is $\sigma$. Then, under the approximation that steps in $\delta$ as the window scale is varied are uncorrelated, $\diff f/\diff\ln\sigma$ is predicted to have the same form for any power spectrum.\footnote{Universality with respect to redshift has also been explored and is not necessarily a prediction of excursion set theory with uncorrelated steps, because the threshold $\dc$ can be allowed to vary with time \citep[e.g.][]{1993MNRAS.262..627L}.}
This universality arises because
\begin{equation}
    \left\langle\delta_a\delta_b\right\rangle 
    = \min\{\sigma_a,\sigma_b\}^2
\end{equation}
if steps in $\delta$ are uncorrelated, irrespective of the power spectrum.
Much attention has been paid to whether the mass functions in simulations are indeed universal and the degree to which they deviate \citep{2001MNRAS.321..372J,2002ApJS..143..241W,2003MNRAS.346..565R,2006ApJ...646..881W,2007ApJ...671.1160L,2007MNRAS.374....2R,2008ApJ...688..709T,2010MNRAS.403.1353C,2011ApJ...732..122B,2011ApJS..195....4M,2011MNRAS.410.1911C,2013MNRAS.433.1230W,2014MNRAS.439.3156J,2016MNRAS.456.2361B,2016MNRAS.456.2486D,2020ApJ...901....5B,2020ApJ...903...87D,2022MNRAS.509.6077O}.

\begin{figure}
	\centering
	\includegraphics[width=\columnwidth]{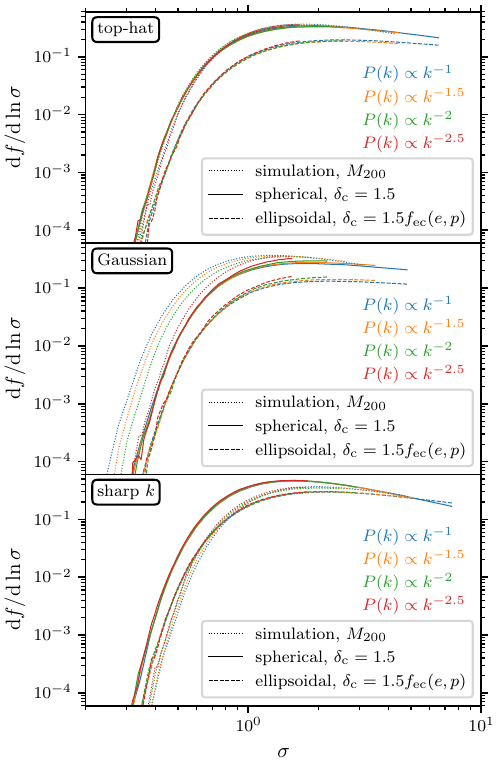}
	\caption{Mass functions expressed in terms of the rms variance $\sigma$, which is set by the mass scale (see Eq.~\ref{sigma}). Specifically, we show the differential fraction $\diff f/\diff\ln\sigma$ of mass in haloes on whose mass scale the rms variance in $\delta$ is $\sigma$. Solid and dashed curves show excursion set predictions for the constant $\dc=1.5$ threshold and for an ellipsoidal-collapse-motivated $\dc=1.5\fec(e,p)$ threshold, respectively, while the dotted curves show simulation results for the $M_{200}$ mass definition. Different colours use different power spectra $P(k)$, while different panels correspond to different window functions. Note that the simulation results depend on the window function because it sets how $\sigma$ is related to the halo mass $M$.
 For the sharp $k$-space window, excursion set predictions are manifestly independent of $P(k)$. For the other windows, this is nearly, but not exactly, the case. For both top-hat and sharp-$k$ windows, the simulation results are also nearly independent of the power spectrum.
 }
	\label{fig:universal}
\end{figure}

However, excursion set theory with properly correlated steps does not predict that $\diff f/\diff\ln\sigma$ is necessarily universal.
For example, for a Gaussian window and a $P(k)\propto k^n$ power spectrum,
\begin{equation}
    \left\langle\delta_a\delta_b\right\rangle 
    = \left[\left(\sigma_a^{-4/(n+3)}+\sigma_b^{-4/(n+3)}\right)/2\right]^{-(n+3)/2},
\end{equation}
which manifestly depends on the spectral index $n$.
In this section, we explore the universality of our numerically evaluated $\diff f/\diff\ln\sigma$. 

\subsection{Scale-free cosmology}

Figure~\ref{fig:universal} shows
\begin{equation}
    \frac{\diff f}{\diff\ln\sigma}=\frac{\diff f}{\diff\ln M}\left|\frac{\diff\ln\sigma}{\diff\ln M}\right|^{-1}
\end{equation}
as a function of $\sigma$ for the mass functions $\diff f/\diff\ln M$ considered in Secs. \ref{sec:massfunctions} and~\ref{sec:selfsim}, where $\sigma$ is a function of the windowing mass scale $M$ as in Eq.~(\ref{sigma}).
The solid curves show the excursion set predictions with $\dc=1.5$, the modified spherical collapse threshold that we showed in Fig.~\ref{fig:hmf_tophat} to accurately predict simulated mass functions when a top-hat window is used. We also show predictions with a similarly rescaled ellipsoidal collapse threshold $\dc=1.5\fec(e,p)$ (dashed curves). Different panels consider different window functions. The dotted curves show the simulation mass functions from Sec.~\ref{sec:simulation}; note that these depend on the window function because $\sigma$ does.

With the sharp $k$-space window (bottom panel), the excursion set $\diff f/\diff\ln\sigma$ are the same for all four power spectra (different colours), as they must be, since steps in $\delta$ are uncorrelated in this case. The simulation mass functions for this window function are also nearly universal and are close to the $\dc=1.5\fec(e,p)$ predictions, although there are systematic differences.
For the Gaussian window (middle panel), the excursion set predictions are close to universal, but small but noticeable deviations are apparent. However, the simulation mass functions are far from universal with this window, which indicates that even further rescaling of the threshold $\dc$ (which shifts the predicted $\diff f/\diff\ln\sigma$ uniformly to the left or right) has no hope of yielding excursion set predictions that match all of the simulations.

\begin{figure}
	\centering
	\includegraphics[width=\columnwidth]{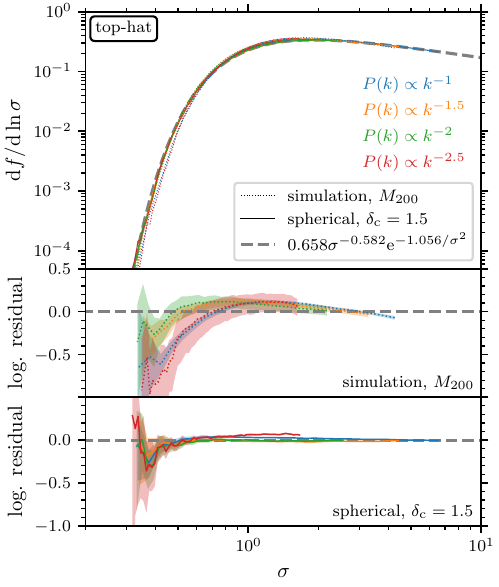}
	\caption{Mass function $\diff f/\diff\ln\sigma$ (as in Fig.~\ref{fig:universal}) for the top-hat window function, comparing excursion set predictions with $\dc=1.5$ (solid curves) and simulation results with the $M_{200}$ mass definition (dotted curves) to the function in Eq.~(\ref{fit}) (thick dashed curve). Different colours correspond to different power spectra. In the lower panels, we show the (base $\e$) logarithmic residuals $\Delta\ln(\diff f/\diff\ln\sigma)$ from this function.
	These panels also include 90 per cent confidence uncertainty bands.
	}
	\label{fig:universal_residuals}
\end{figure}

Interestingly, for the top-hat window (top panel), the excursion set $\diff f/\diff\ln\sigma$ appear to be almost exactly universal, at least over the range of power spectra and $\sigma$ considered here. The simulation mass functions are also universal\footnote{\citet{2020ApJ...903...87D} found that mass functions in the same simulations deviate from universality to a greater degree. This is because \citet{2020ApJ...903...87D} evaluated $\sigma$ using power spectra modified to account for the simulation box sizes. We discuss the impact of the box size in different terms (Sec.~\ref{sec:conditional}).} and match the $\dc=1.5$ predictions, as required by Fig.~\ref{fig:hmf_tophat}. Figure~\ref{fig:universal_residuals} shows that for the top-hat window, the $\dc=1.5$ predictions and the simulation mass functions (with $M_{200}$ mass definition) are closely approximated by the function
\begin{equation}\label{fit}
    \frac{\diff f}{\diff\ln\sigma}=0.658\sigma^{-0.582}\e^{-1.056/\sigma^2}.
\end{equation}
In the lower panel of Fig.~\ref{fig:universal_residuals}, we show the residuals from this function, which only start to exceed 10 per cent in the very high-mass (low-$\sigma$) tail of the distribution. In contrast to the most common parametrization of halo mass functions \citep[e.g.][]{2008ApJ...688..709T}, note that there is no additive constant in the prefactor to the exponential in Eq.~(\ref{fit}). We also constrain this function to integrate to 1, since excursion set theory with a constant threshold must eventually associate every particle with a halo if $\sigma$ can become arbitrarily large.

\begin{figure}
	\centering
	\includegraphics[width=\columnwidth]{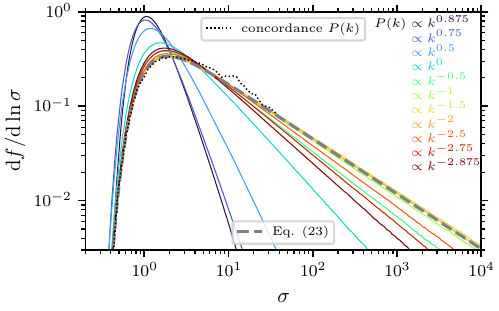}
	\caption{Excursion set mass functions $\diff f/\diff\ln\sigma$ (as in Fig.~\ref{fig:universal}) for the top-hat window function with constant threshold $\dc=1.5$. We consider a wide range of scale-free power spectra (different colours) and $\sigma$. The apparent universality in Fig.~\ref{fig:universal_residuals}, indicated by the dashed line here (Eq.~\ref{fit}), evidently only applied to a narrow range of parameters. The thin dotted curve shows the $\diff f/\diff\ln\sigma$ predicted for a concordance cosmology power spectrum at redshift $z=0$ (Fig.~\ref{fig:concordance_power}). It nevertheless matches the ``universal'' curve (Eq.~\ref{fit}; dashed curve) reasonably well, except near $\sigma\sim 10$-20, a deviation that is linked to the presence of a feature in the power spectrum (see Figs. \ref{fig:concordance_power} and~\ref{fig:concordance}).}
	\label{fig:universal_ext}
\end{figure}

However, the range of $\sigma$ and power spectra represented in Fig.~\ref{fig:universal_residuals} is narrow. 
For the same choice of a top-hat window with $\dc=1.5$, Fig.~\ref{fig:universal_ext} shows $\diff f/\diff\ln\sigma$ evaluated for a much wider range of $\sigma$ and for $P(k)\propto k^n$ with $n$ ranging from $-2.875$ up to $0.875$.\footnote{$\sigma$ diverges for $n\leq-3$ due to the influence of large scales, and for the top-hat window, $\sigma$ diverges for $n\geq 1$ due to the influence of small scales.}
Here we discretize the window scales such that successive $\sigma$ are separated by $\Delta\ln\sigma=0.002$, we sample $10^5$ trajectories in $\delta$ for each power spectrum, and we count first-crossing $\sigma$s in bins of width $\Delta\ln\sigma=0.05$. As in Sec.~\ref{sec:massfunctions}, we stack the distributions obtained for different growth factors $D$ separated by factors of 1.03. It is clear in Fig.~\ref{fig:universal_ext} that $\diff f/\diff\ln\sigma$ is only nearly universal for $P(k)\propto k^n$ with $-2\lesssim n\lesssim -1$; in these cases $\diff f/\diff\ln\sigma$ closely matches Eq.~(\ref{fit}) (dashed curve). Large deviations arise as $n$ approaches $-3$, and even larger deviations arise as $n$ approaches $1$.\footnote{Note that $n=0$ corresponds to Poisson noise, so steps in the top-hat-windowed $\delta$ are uncorrelated. By the argument of \citet{1991ApJ...379..440B}, the exact mass function is $\diff f/\diff\ln \sigma=\sqrt{2/\pi}(\dc/\sigma)\e^{-\dc^2/(2\sigma^2)}$ in this case.}

\begin{figure}
	\centering
	\includegraphics[width=\columnwidth]{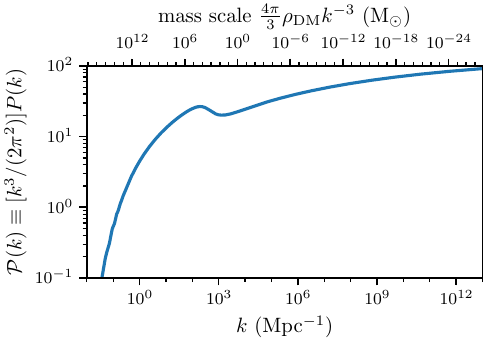}
	\caption{Cold dark matter power spectrum at redshift $z=0$ for \citet{2020A&A...641A...6P} cosmological parameters, as extrapolated at linear order using cosmological perturbation theory. Growth of dark matter perturbations for $k\gtrsim 10^2$~Mpc$^{-1}$ is suppressed because baryonic matter resists clustering on such small scales. On the top, we show the approximate mass scale associated with each wavenumber $k$.}
	\label{fig:concordance_power}
\end{figure}

\subsection{Spectral features}

Moreover, realistic cosmologies are not scale-free. The dotted curve in Fig.~\ref{fig:universal_ext} shows the excursion set prediction for the concordance cosmology power spectrum at redshift $z=0$. This power spectrum, shown in Fig.~\ref{fig:concordance_power}, is evaluated using cosmological perturbation theory at linear order using the \textsc{CLASS} Boltzmann solver \citep{2011JCAP...07..034B}, and we extrapolate below the code's resolution limit with the analytic solution of \citet{1996ApJ...471..542H}. We adopt cosmological parameters from \citet{2020A&A...641A...6P}.\footnote{We do not impose a small-scale cutoff to the power spectrum, which depends on dark matter microphysics \citep[e.g.][]{2004MNRAS.353L..23G}. Excursion set theory with a top-hat window improperly predicts haloes of arbitrarily low mass even when there is a small-scale cutoff \citep[e.g.][]{2013MNRAS.428.1774B}, and we have verified that properly correlated steps do not cure this defect.}
We use the natural normalization for the top-hat window mass, $M=(4\pi/3)\rho_\mathrm{DM}r^3$, where $\rho_\mathrm{DM}\simeq 3.31\times 10^{10}$~M$_\odot$Mpc$^{-3}$ is the comoving dark matter density.
We generate $10^5$ trajectories in $\delta$ using window masses ranging from $10^{-30}$~M$_\odot$ (for which $\sigma\simeq 40$) to $1.1\times 10^{16}$~M$_\odot$ (for which $\sigma\simeq 0.28$) and separated by $\Delta\ln \sigma = 0.002$. We count first-crossing $\sigma$s in bins of width $\Delta\ln\sigma=0.1$.

\begin{figure}
	\centering
	\includegraphics[width=\columnwidth]{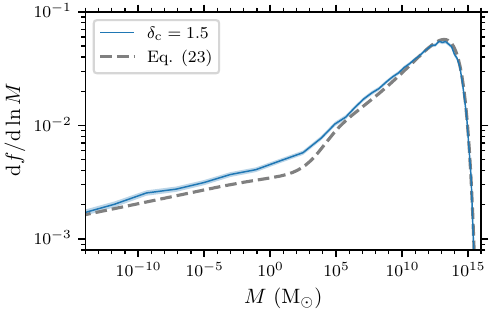}
	\caption{Differential fraction $\diff f/\diff\ln M$ of dark matter in haloes of mass $M$ for a concordance cosmology at redshift $z=0$; we use the power spectrum in Fig.~\ref{fig:concordance_power}. The solid blue curve shows the excursion set prediction for a top-hat window and $\dc=1.5$ threshold, while the shading indicates the $2\sigma$ Poisson uncertainty range. The dashed curve shows the mass function predicted by Eq.~(\ref{fit}).
    Equation~(\ref{fit}) matches the excursion set prediction reasonably well except near the dip in halo abundance below around $10^3$~M$_\odot$, which is associated with a feature in the power spectrum (see Fig.~\ref{fig:concordance_power}).
    }
	\label{fig:concordance}
\end{figure}

For this power spectrum, the spectral index $\diff\ln P/\diff\ln k$ runs from close to $1$ at large scales down to nearly $-3$ at small scales, but this effect does not lead the predicted $\diff f/\diff\ln\sigma$ (dotted curve in Fig.~\ref{fig:universal_ext}) to deviate significantly from Eq.~(\ref{fit}). Instead, there is only significant deviation near $\sigma\sim 10$-20. This is associated with a characteristic feature in the power spectrum (Fig.~\ref{fig:concordance_power}) near $k\sim 10^{2.5}$~Mpc$^{-1}$, which arises because baryonic matter resists clustering at smaller scales, suppressing the growth rate of dark matter perturbations on those scales \citep[e.g.][]{1996ApJ...471..542H,2006PhRvD..74f3509B}.
The effect of this deviation from universality is clearer in Fig.~\ref{fig:concordance}, which shows the halo mass function $\diff f/\diff\ln M$ in terms of mass $M$. Equation~(\ref{fit}) (dashed curve) predicts a sharp drop in halo abundance below about $10^3$~M$_\odot$. The excursion set theory with $\dc=1.5$ (solid black curve with Poisson error band) shallows that dip.

Nevertheless, the mass functions predicted by the $\dc=1.5$ model are reasonably universal, and Eq.~(\ref{fit}) works very well, for the $\sigma<10$ and spectral indices $n\equiv\diff\ln P/\diff\ln k<-1$ that are relevant to the haloes of galaxies and galaxy clusters. It is useful to rewrite Eq.~(\ref{fit}) as
\begin{equation}\label{fit2}
    \frac{\diff f}{\diff\ln\nu}=0.519\nu^{0.582}\e^{-0.469\nu^2},
\end{equation}
where we define
\begin{equation}
    \nu\equiv \dc/\sigma.
\end{equation}
Note that the differential mass fraction itself is then
\begin{equation}
    \frac{\diff f}{\diff\ln M}
    =
    \left.\frac{\diff f}{\diff\ln\nu}\right|_{\nu=\dc/\sigma_M}\,
    \left|\frac{\diff\ln\sigma_M}{\diff\ln M}\right|,
\end{equation}
since $\diff\ln\nu/\diff\ln\sigma=-1$, and the differential halo number density is related by Eq.~(\ref{dndlogM}).
Not only is Eq.~(\ref{fit2}) valid for arbitrary $\dc$, but it can be applied to conditional mass functions as well, as we discuss next.

\subsection{Conditional mass functions}

Focusing on trajectories that are constrained to cross $\delta_0$ at $\sigma_0$, we may redefine
\begin{equation}\label{nucond}
    \nu\equiv(\dc-\delta_0)/(\sigma^2-\sigma_0^2)^{1/2}
\end{equation}
and let $\diff f/\diff\ln\nu$ be the differential fraction of these trajectories that first cross $\dc$ at $\nu$ (which depends on $\sigma$).
Since $\nu$ is a function of $\sigma$, which is in turn a function of the mass scale, $\diff f/\diff\ln\nu$ is a conditional mass function of the sort discussed in Sec.~\ref{sec:conditional}, useful for studying halo clustering bias and the mass functions of halo progenitors. For this definition,
\begin{equation}
    \frac{\diff f}{\diff\ln M}
    =
    \left.
    \frac{\diff f}{\diff\ln\nu}\right
    |_{\nu=\frac{\dc-\delta_0}{(\sigma_M^2-\sigma_0^2)^{1/2}}}\,
    \left|
    \frac{\sigma_M^2}{\sigma_M^2-\sigma_0^2}\frac{\diff\ln\sigma_M}{\diff\ln M}
    \right|
\end{equation}
is the differential fraction, in haloes of mass $M$, of particles that satisfy the condition.
For excursion set theory with uncorrelated steps and a constant threshold, $\diff f/\diff\ln\nu$ necessarily has the same form, not only for any power spectrum, but also for any $\delta_0$ and $\sigma_0$. However, with properly correlated steps, $\diff f/\diff\ln\nu$ may depend on all of these parameters.

In Appendix~\ref{sec:conditionaluniversal}, we test the universality of $\diff f/\diff\ln\nu$, interpreted as a conditional mass function as described above, for excursion set theory with top-hat windowing. We show that $\diff f/\diff\ln\nu$ is approximately universal, and is fit well by Eq.~(\ref{fit2}), only as long as $\dc\gtrsim\delta_0+\sigma_0$. This condition would often be satisfied in the context of halo clustering bias, where the region under consideration is much larger than the haloes (so $\sigma\gg\sigma_0$). However, it would fail, for example, when considering progenitor mass functions at a recent time (so $\dc$ is not much larger than $\delta_0$), which are relevant when constructing halo merger trees. In the opposite regime, $\dc\lesssim\delta_0+\sigma_0$, the mass function $\diff f/\diff\ln\nu$ depends strongly on $\delta_0$ and $\sigma_0$, and it also depends on whether or not the crossing of $\delta_0$ at $\sigma_0$ is constrained to be the first crossing of $\delta_0$. The first-crossing constraint is appropriate for progenitor mass functions but is not appropriate for studies of halo clustering bias. Compared to Eq.~(\ref{fit2}), $\diff f/\diff\ln\nu$ tends to become more sharply peaked in the $\dc\lesssim\delta_0+\sigma_0$ regime, and it peaks at lower values of $\nu$ (lower masses).

\section{Conclusions}\label{sec:conclusion}

We evaluated halo mass functions from excursion set theory by direct numerical sampling, without the simplifying approximations used in most previous work.
When a real-space spherical top-hat window function is employed, excursion set theory with a constant $\dc=1.5$ threshold accurately predicts halo mass functions with the $M_{200}$ mass definition in cosmological simulations of a range of matter-dominated cosmologies.
The model is also able to account for halo clustering bias and to predict progenitor mass functions with good accuracy.
For a concordance $\Lambda$CDM cosmology, predicted mass functions lie between simulated mass functions with the $M_\mathrm{200m}$ and $M_\mathrm{200c}$ mass definitions once dark energy starts to become important.

In contrast, a nonconstant threshold based on ellipsoidal collapse predicts too few haloes of every mass, except if a sharp $k$-space window function is used.
The physical reason for this surprising outcome is unclear, but we note that the physical picture relevant to excursion set theory is significantly more complicated than that assumed by the modeling that motivates the ellipsoidal collapse threshold.
Moreover, our result is consistent with previous studies that used machine learning to explore which information in the initial conditions is needed to predict halo masses.
Apparently, the simplest excursion set theory -- that with the standard spherical top-hat window and a constant threshold -- is sufficient to produce accurate halo mass functions and assembly histories. The widely discussed refinement of considering a nonconstant threshold does not appear to be needed or appropriate.

Excursion set theory with the top-hat window and $\dc=1.5$ threshold predicts a halo mass function that is nearly universal when expressed in terms of the rms density variance $\sigma$ and is closely approximated by Eq.~(\ref{fit2}). In this sense, it agrees with a wide variety of studies noting that the mass function is either universal or nearly so, although the functional form of Eq.~(\ref{fit2}) is somewhat different from what has been previously proposed.
Moreover, within the regimes relevant to studies of halo clustering bias or of halo progenitors in the distant past, the accuracy of Eq.~(\ref{fit2}) even extends to conditional mass functions.
However, significant deviations from universality can arise at very small mass scales, for extreme spectral indices, or when there are features in the power spectrum.
Conditional mass functions greatly deviate from universality in the range of parameters relevant to studies of recent progenitors (so, for example, the approximation of Eq.~\ref{fit2} is likely less suitable for constructing merger trees than is the direct calculation). It remains to be seen whether all of these deviations would accord with simulation results.

\section*{Acknowledgements}

The author thanks Simon White for helpful discussions and advice and Andrew Benson, Benedikt Diemer, and Ethan Nadler for comments on the manuscript.

%%%%%%%%%%%%%%%%%%%%%%%%%%%%%%%%%%%%%%%%%%%%%%%%%%
\section*{Data Availability}
 
Halo catalogues for the simulations of \citet{2015ApJ...799..108D} used in this work are publicly available as detailed by \citet{2020ApJS..251...17D}.



\bibliographystyle{mnras}
\bibliography{main}

%%%%%%%%%%%%%%%%%%%%%%%%%%%%%%%%%%%%%%%%%%%%%%%%%%

%%%%%%%%%%%%%%%%% APPENDICES %%%%%%%%%%%%%%%%%%%%%

\appendix

\section{Convergence of excursion set mass functions}\label{sec:convergence}

\begin{figure*}
	\centering
	\includegraphics[width=\linewidth]{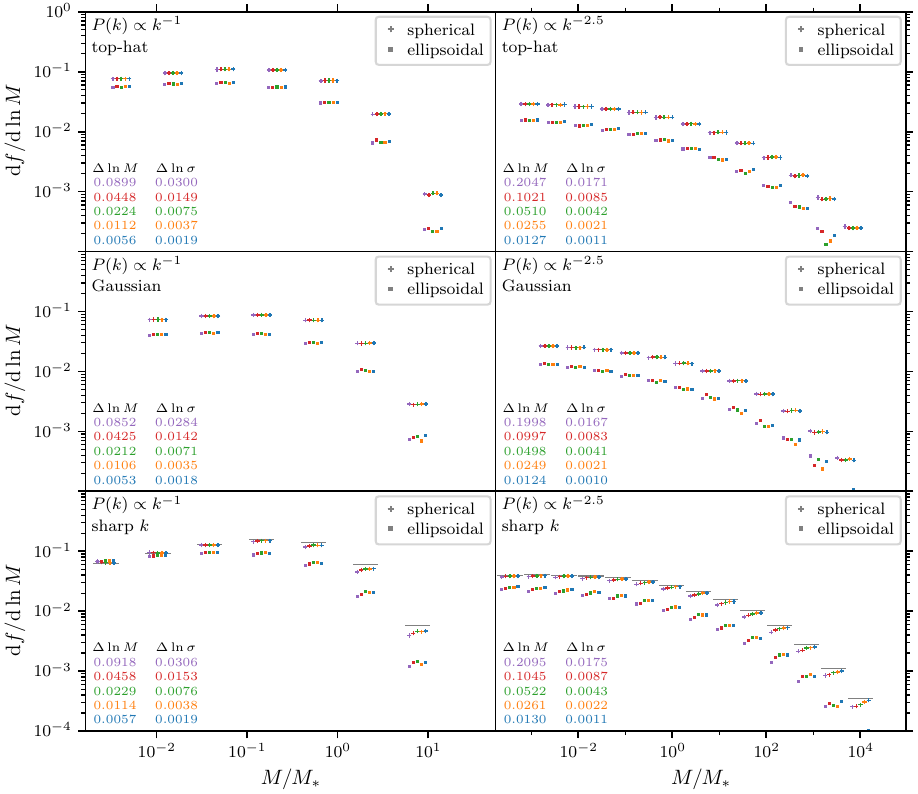}
	\caption{Testing the degree to which numerically sampled excursion set halo mass functions are converged. As in Sec.~\ref{sec:massfunctions}, we plot the differential fraction $\diff f/\diff\ln M$ of particles deemed to reside in haloes of mass $M$, considering both the spherical collapse threshold $\dc=1.686$ (crosses) and the ellipsoidal collapse threshold $\dc=1.686\fec(e,p)$ (squares).
    Each sequence of 5 differently coloured points marks $\diff f/\diff\ln M$ at the same mass $M$; the horizontal separation is only for visual convenience. We evaluate these $\diff f/\diff\ln M$ for different numbers $N$ of logarithmically spaced mass scales: $N=200$ (purple), 400 (red), 800 (green), 1600 (orange), and 3200 (blue). The corresponding steps in $\ln M$ and $\ln\sigma$ are indicated in the respective panels.
    The left-hand and right-hand panels consider power spectra $P(k)\propto k^n$ with $n=-1$ and $-2.5$, respectively. The upper, central, and lower panels use top-hat, Gaussian, and sharp $k$-space windows, respectively.
    For spherical collapse, we generate these $\diff f/\diff\ln M$ using a sample of $10^5$ trajectories in $\delta$, while for ellipsoidal collapse, we use only $10^4$ trajectories in $T_{ij}$, which leads to a higher level of statistical noise.
    For the sharp-$k$ window (bottom) with the spherical collapse threshold (crosses), we mark the exact analytic $\diff f/\diff\ln M$ with horizontal lines.
    For this window, it is difficult to achieve convergence even for very high $N$, a consequence of how abruptly the trajectories in $\delta$ can jump (see Fig.~\ref{fig:example_excursions}).
    However, for the top-hat and Gaussian windows, the trajectories in $\delta$ are much smoother, and the $\diff f/\diff\ln M$ appear to be reasonably well converged as long as $\Delta\ln \sigma\lesssim 0.01$.
    }
	\label{fig:convergence}
\end{figure*}

\begin{figure*}
	\centering
	\includegraphics[width=\linewidth]{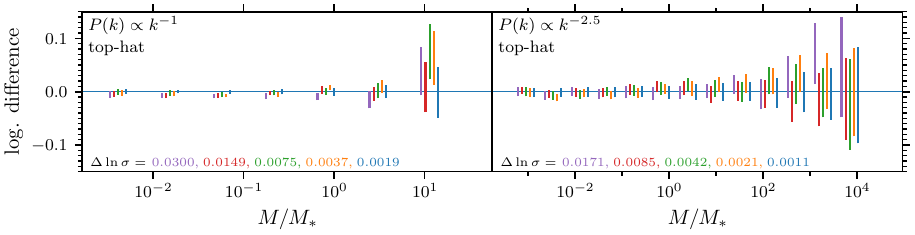}
	\caption{
	A closer look at numerical convergence of excursion set predictions with top-hat window and constant threshold $\dc=1.686$. Here we show the (base $\e$) logarithmic difference $\Delta\ln(\diff f/\diff\ln M)$ between the mass functions evaluated with $N\leq 3200$ mass scales (different colours) and those evaluated with $N=3200$ mass scales (blue). We show 90 per cent confidence uncertainty bars. As in Fig.~\ref{fig:convergence}, the two panels correspond to different power spectra, and different colours (different $N$) are horizontally offset for visual convenience only.
	}
	\label{fig:convergence2}
\end{figure*}

In our numerical approach to evaluating excursion set mass functions (Secs. \ref{sec:sampling} and~\ref{sec:massfunctions}), we discretize the windowing scales $r_1<r_2<...<r_N$ and search for the ``first crossing'', the largest $r_i$ for which $\delta_i$ exceeds the threshold $\dc$. It is possible that the first crossing actually occurred at a larger $r$, but that $\delta>\dc$ for only a narrow range of scales that did not include any of the $r_i$. In this appendix, we test the degree to which this effect might impact our results. Specifically, we explore how the mass function $\diff f/\diff\ln M$ depends on how finely the windowing scales $r_i$ are spaced.

In Sec.~\ref{sec:massfunctions}, we used $N$ window scales between the smallest scale (for which $\sigma$ ranged from about 30 to about 400, depending on the power spectrum) and the largest scale (for which $\sigma=0.28$), where $N=800$ for the six-dimensional $T_{ij}$ trajectories and $N=3200$ for the one-dimensional trajectories in $\delta$.
Here, we vary $N$ from 200 to 3200 for both $T_{ij}$ and $\delta$. For each value of $N$, we sample $10^5$ trajectories in $\delta$, but to spare computational expense, we sample only $10^4$ trajectories in $T_{ij}$. To suppress statistical variance, we count first-crossing masses in bins of increased width $\Delta\ln M = 1.3$. As in Sec.~\ref{sec:massfunctions}, we further suppress statistical variance by stacking the first-crossing distributions for different growth factors $D$.

We consider the $P(k)\propto k^n$ power spectra with $n=-1$ and $-2.5$ and each of the three window functions. For five values of $N$ ranging from 200 to 3200, Fig.~\ref{fig:convergence} shows the mass functions $\diff f/\diff\ln M$ that result from this evaluation. For the sharp $k$-space window (lower panels), the horizontal lines show the analytic mass function of \citet{1974ApJ...187..425P}, which is exact in this case. To allow for an accurate comparison with the excursion set $\diff f/\diff\ln M$ in such wide bins, we integrate this analytic mass function over the width of each bin and then divide it by the bin width.

Evidently, convergence is difficult to achieve for the sharp $k$-space window (bottom). $\diff f/\diff\ln M$ depends significantly in this case on the number $N$ of windowing scales. Even at the highest $N$ (blue), the excursion set method predicts too few high-mass haloes and (for $P(k)\propto k^{-1}$) too many low-mass haloes, exactly as would be expected if some threshold crossings are being missed. This difficulty is a consequence of how abruptly the trajectories in $\delta$ for this window can jump, as exemplified in Fig.~\ref{fig:example_excursions}.

However, the trajectories in Fig.~\ref{fig:example_excursions} are much smoother for the top-hat and Gaussian windows, so we do not expect convergence to be as difficult to achieve in these cases. Indeed, in the upper and central panels of Fig.~\ref{fig:convergence}, $\diff f/\diff\ln M$ exhibits much less variation with the number $N$ of windowing scales. Although exact expressions for $\diff f/\diff\ln M$ are not known in these cases, convergence with respect to the discretization scheme appears to be achieved when the spacing of successive window scales corresponds to $\Delta\ln \sigma\lesssim 0.01$.
Since excursion set predictions with top-hat window and constant threshold are central to this work, Fig.~\ref{fig:convergence2} presents a closer look at convergence in this case.

\section{Universality of conditional mass functions}\label{sec:conditionaluniversal}

\begin{figure*}
	\centering
	\includegraphics[width=\linewidth]{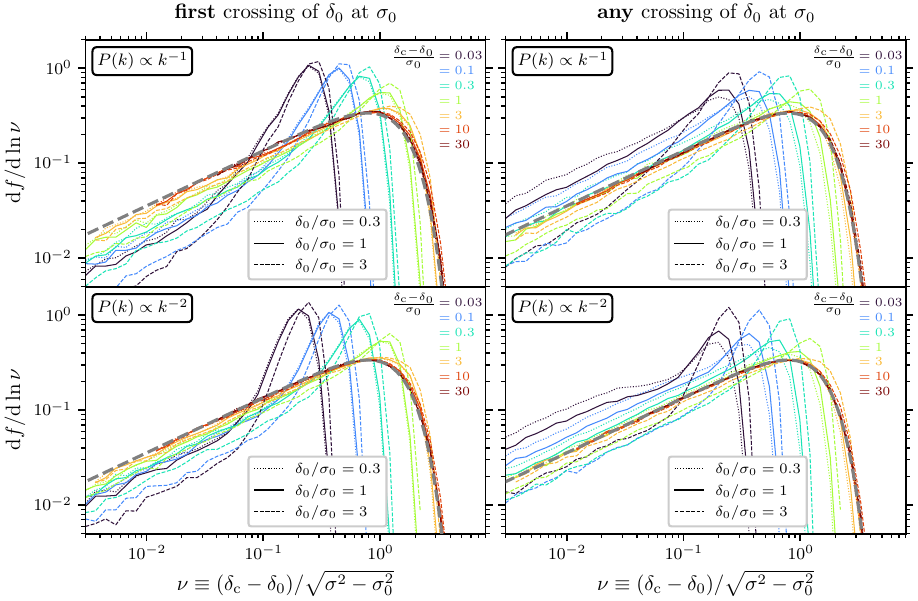}
	\caption{Conditional mass functions expressed in terms of $\nu\equiv(\dc-\delta_0)/(\sigma^2-\sigma_0^2)^{1/2}$, where $\sigma$ relates to the mass scale in accordance with Eq.~(\ref{sigma}). Specifically, we plot the differential fraction of trajectories that first cross the threshold $\dc$ at $\sigma$, given a crossing of $\delta_0$ at $\sigma_0$. In the left-hand panels, the crossing of $\delta_0$ at $\sigma_0$ is further constrained to be the first such crossing, while in the right-hand panels, it is not. The upper panels adopt $P(k)\propto k^{-1}$, while the lower panels adopt $P(k)\propto k^{-2}$. We consider a range of $\delta_0/\sigma_0$ and $\dc/\sigma_0$, indicated by the different line colours and styles. As long as $(\dc-\delta_0)/\sigma_0\gtrsim 1$ (yellow through brown), the conditional mass functions closely match Eq.~(\ref{fit2}), represented by the thick dashed gray curve. However, when $(\dc-\delta_0)/\sigma_0\lesssim 1$, the mass functions begin to vary greatly as a function of all relevant parameters.
    }
	\label{fig:conditionaluniversal}
\end{figure*}

In this appendix, we test the degree to which conditional mass functions are universal, when predicted by excursion set theory with a constant threshold and top-hat windowing. We consider the power spectra $P(k)\propto k^n$ with $n=-1$ and $n=-2$, a choice motivated by the observation in Sec.~\ref{sec:universal} that unconditional mass functions for $-2\lesssim n\lesssim -1$ are nearly universal and closely match Eq.~(\ref{fit2}).
We use the method of Sec.~\ref{sec:conditionalsampling} to sample trajectories that cross $\delta_0$ at the mass scale for which the rms variance is $\sigma_0$. For $\sigma<\sigma_0$, we use window scales separated by $\Delta\ln\sigma=0.002$ (down to $\sigma=0.1\sigma_0$), while for $\sigma>\sigma_0$, we separate the window scales by $\Delta\ln\sqrt{\sigma^2-\sigma_0^2}=0.002$ (down to $\sqrt{\sigma^2-\sigma_0^2}=10^{-3}$).

Figure~\ref{fig:conditionaluniversal} shows the resulting conditional mass functions, expressed in terms of $\nu\equiv(\dc-\delta_0)/(\sigma^2-\sigma_0^2)^{1/2}$. To produce these, we count first crossings of $\dc$ in bins of width $\Delta\ln\nu=0.2$. We consider a range of $\delta_0/\sigma_0$ and $\dc/\sigma_0$, and we separately consider trajectories that first cross $\delta_0$ at $\sigma_0$ (left-hand panels) and trajectories for which a first crossing is not demanded (right-hand panels). Each curve is produced from a sample of $10^5$ trajectories.
For comparison, the thick dashed curve shows Eq.~(\ref{fit2}).

We find that the form of the mass function $\diff f/\diff\ln\nu$ depends most significantly on $(\dc-\delta_0)/\sigma_0$ (different colours). When $(\dc-\delta_0)/\sigma_0\gtrsim 1$, the mass functions agree well with the nearly universal form of Eq.~(\ref{fit2}). However, when $(\dc-\delta_0)/\sigma_0\lesssim 1$, mass functions depend strongly on $(\dc-\delta_0)/\sigma_0$, becoming more sharply peaked and peaking at lower $\nu$ as $(\dc-\delta_0)/\sigma_0$ decreases. In this regime, $\diff f/\diff\ln\nu$ also depends on $\delta_0/\sigma_0$ (different line styles), on the power spectrum, and on whether the crossing of $\delta_0$ at $\sigma_0$ is constrained to be the first crossing. Note that the first-crossing constraint is appropriate when studying progenitor mass functions but inappropriate when studying halo clustering bias.

%%%%%%%%%%%%%%%%%%%%%%%%%%%%%%%%%%%%%%%%%%%%%%%%%%

% Don't change these lines
%\bsp	% typesetting comment
\label{lastpage}
\end{document}